\documentclass[12pt,preprint]{aastex}

\received{TBD}
\begin{document}

\title{Systematic and quantitative approach for the identification of \\
high energy $\gamma$-ray source populations}

\author{Diego F. Torres$^1$ and Olaf Reimer$^{2}$}

\altaffiltext{1}{Lawrence Livermore National Laboratory, 7000 East
Ave., L-413, Livermore, CA 94550, E-mail: dtorres@igpp.ucllnl.org}
\altaffiltext{2}{Institut f\"ur Theoretische Physik IV,
Ruhr-Universit\"at Bochum, 44780, Germany. E-mail:
olr@tp4.ruhr-uni-bochum.de Now at: W. W. Hansen
Experimental Physics Laboratory, Stanford University, Stanford, CA
94305}

\begin{abstract}

A large fraction of the detections to be made by the Gamma-ray
Large Area Space Telescope (GLAST) will initially be unidentified.
We argue that traditional methodological approaches to identify
individuals and/or populations of $\gamma$-ray sources will
encounter procedural limitations. These limitations will hamper
our ability to classify source populations lying in the
anticipated dataset with the required degree of confidence,
particularly those for which no member has yet been convincingly
detected in the predecessor experiment EGRET. Here we suggest a
new paradigm for achieving the classification of $\gamma$-ray
source populations based on the implementation of an a priori
protocol to search for theoretically-motivated candidate sources.
In order to protect the discovery potential of the sample, it is
essential that such paradigm will be defined before the data is
unblinded. Key to the new procedure is a statistical assessment by
which the discovery of a new population can be claimed.  Although
we explicitly refer here to the case of GLAST, the scheme we
present may be adapted to other experiments confronted with a
similar problematic.

\end{abstract}


\section{Problem statement}

The anticipated source wealth from observations carried out by the
upcoming $\gamma$-ray mission GLAST, potentially yielding the
discovery of thousands of new high-energy sources following
extrapolations from predecessor experiments, will create several
problems for source identification.\footnote{For general
description of the GLAST main instrument LAT and its potential see
the GLAST Science Requirements Document 2003.} Catalogs of the
most likely candidate sources (Active Galactic Nuclei -AGNs-, and
neutron stars/pulsars -PSRs- ) will very likely not be complete to
the required low radio and/or X-ray flux levels required for
counterpart studies, predictably leaving many of these
$\gamma$-ray detections unidentified. And even if the pulsar and
AGN catalogs were sufficiently deep, they may not yield
unambiguous source identifications: A complete catalog for the
anticipated numbers of sources, projected using the instrumental
point spread function (psf), would generate total sky coverage,
with one or more candidates in every line-of-sight for incident
photons corresponding to their (energy dependent) psf.
This would hamper or even prevent unambiguous source
identifications based solely on positional correlation. In
addition, a legacy from the EGRET experiment is the indication
that we are already failing to identify one or more new source
populations with astrophysical objects, both at low and at high
Galactic latitude. Specifically, the identification of variable,
non-periodic, point-like sources at low galactic latitude, as well
as of non-variable sources at high latitude is still missing
(Reimer 2001), all those exhibiting characteristics not found
among the EGRET-detected pulsars or blazars.




In the LAT era and beyond, if it is the objective to conclusively
identify all individual $\gamma$-ray source detections, we will
predictably fail. The anticipated number of counterparts, their
relative faintness deduced from luminosity functions, the missing
all-sky coverage in the relevant wavebands for deep counterpart
studies, and the expected ambiguities due to source confusion in
densely populated regions of the $\gamma$-ray sky will preclude
reaching this ultimate goal of source identification.
Consequently, we should aim to identify at least all {\it classes}
of sources, and subsequently attempt to gain in--depth
astrophysical knowledge by studying the most interesting
representatives among such populations. Let us deepen into these
statements in the following.

The anticipated number of unidentified detections will preclude
the making of an individual deep multifrequency study for every
source, in the way it led to the identification of many
$\gamma$-ray blazars and the Geminga pulsar.

Suppose that we have a sufficiently complete counterpart catalog,
such that a member of it spatially coincides with most of the LAT
sources. Does this imply that we have already identified all
sources? To answer this question consider that we have, instead, a
reasonably complete sky coverage of sources, i.e. GRBs as an
example. An overlay of all error boxes of GRBs reported from BATSE
covers the whole sky. Then, there is at least one GRB spatially
coinciding with any possible counterpart or host. Consequently,
here a correlation analysis lacks identification capability, even
when it is clear that not all populations of astrophysical objects
are plausible candidates for GRB generation or hosting, nor that
all of them should be probed. More particularly, we can not claim,
using correlation analysis, that GRBs have appeared more often in
starburst or luminous infrared galaxies than in normal galaxies.
Therein lies the dilemma. If the number of unidentified sources
and/or the number of plausible candidates is sufficiently large,
what will constitute an identification? How shall we find evidence
for new populations of sources, and new members within these
populations, in the GLAST era?

The most successful identification scheme for $\gamma$-ray sources
so far is based upon multifrequency follow-up observations,
unless there is a given prediction of flux periodicity, which
itself would unambiguously label the source if resembled in the
data. The latter, however, will happen only for a fraction of LAT
detections, either because of the absence of contemporaneous
pulsar timing solutions, or because of shortage of precise
theoretical predictions for the variability pattern other than
periodicity. Note that variability of $\gamma$-rays probes,
generally, timescales, not periodicities, and can be used
predominantly to {\it rule out} membership into classes and only
when it has been found at a significant level. For example, if a
given source {\it is} variable, we consequently assume that it is
not produced in phenomena on timescales larger than the
corresponding exposure. In essence, this will rule out all
possible counterparts producing steady $\gamma$-ray
fluxes.\footnote{For many of the theoretically anticipated LAT
sources steady $\gamma$-ray emission is predicted. Such candidate
populations are Supernova remnants (e.g., Torres et al. 2003),
luminous infrared galaxies (e.g., Torres et al. 2004), or galaxy
clusters (e.g., Reimer et al. 2003)}

However, if a theoretically compatible variability timescale
exists,
it will prompt the need of carrying out follow-up observations,
which will necessarily require a considerable amount of time and
resources, without guaranteed success of achieving an unique
identification. The bottom line is that adopting this scheme, with
LAT observations, particularly during the first year of data
taking, we may limit our capability to identify new populations of
sources if applying multifrequency follow-up methods only.

If we have a positional correlation between a candidate and an
unidentified $\gamma$-ray source, and in addition there exist a
matching variability timescale between theoretical predictions for
such object and the data, then how can we, with nothing else,
definitely say that an identification was achieved? And even if we
convince ourselves to assert it, how many of such individual cases
should be found in order to claim the discovery of a new
population of sources with satisfactory statistical significance?
How would the latter be quantitatively evaluated? Not having {\it
a priori of the expected number of source detections} a criterion
by which to answer the previous questions will confront us with a
situation of ambiguity between results achieved by applying
different classification standards, with no instance to decide in
a unbiased way whether an identification has been achieved or not.

In order to overcome these predictable problems, a paradigm shift
in the way we seek $\gamma$-ray population classification is
suggested. We need to define a sensitive and quantitative
criterion, by which we could identify both variable and
non-variable populations. This Letter provides a feasible scheme
for defining such a criterion. Although we refer here explicitly
to the case of GLAST (Large Area Telescope, LAT), the scheme we
present can certainly be adapted to other experiments confronted
with a similar combination of problems, for example in
contemporary neutrino astronomy.

\section{Identification of $\gamma$-ray populations}

Here we elaborate a scheme to identify and classify new
$\gamma$-ray source populations.

\subsection{What to search for?}

Starting from a theoretical prediction of a population of
astronomical objects to be detectable above the LAT instrumental
sensitivity, we propose to assume a
\begin{itemize}
\item {\it Theoretical censorship:} we request as part of the
criterion that predictions (ideally of multiwavelength character)
are available for a subset of the proposed class of
counterparts.\footnote{The term {\it predictions} particularly
refers to observable parameters with the given set of instruments
available. For example, predicting polarization may be a truly
sensitive theoretical discriminator, but LAT will not measure
polarization.}
\end{itemize}
This request is made to avoid the blind testing of populations
that may or may not produce $\gamma$-rays, but for which no other
than a spatial correlation result can be achieved a posteriori. If
there is no convincing theoretical support that a population can
emit $\gamma$-rays before conducting the search, such population
may not be sought this way. Although obvious, it should be
explicitly stated that we will not, by applying this method,
disallow the possibility of making serendipity discoveries.
Imposing of a theoretical censorship is not just a matter of
theoretical purity, but rather it is statistically motivated, as
we explain below. Such censorship applies similarly to all {\it a
priori} selection of subclasses, i.e., the imposing of cuts in
samples that are aimed to isolate the members from which we
preferably expect detectable $\gamma$-ray emission. Cut
optimization will certainly have an impact on the budget (a concpt
dicussed below) so an adequate and theoretically motivated
procedure should be defined along with assigning the budget.

\subsection{Protection of discovery potential}

By probing a large number of counterparts candidates with at least
equally large number of trials with the same data set, one {\it
will} find positive correlations, at least as a result of
statistical fluctuations (also referred to as chance
capitalization). Then, to claim significance, one would have to
check if the penalties that must be paid for such a finding (i.e.,
the fact that there were a number of trials that led to null
results) does not overcome the significance achieved. Needless to
say, a number of possible bias are expected to influence the
computation of the penalties. The example here is ultra high
energy cosmic rays (UHECRs), where there are already a number of
dubious, correlation-based, discovery claims, even when the sample
of events is small (see, e.g. Evans et al. 2003, Torres et al.
2003b). GLAST-LAT, and in general $\gamma$-ray astronomy, can
prepare to overcome this difficulty before entering the new era of
photon wealth, as UHECR physics does before unblinding data from
the Pierre Auger observatory (Pierre Auger Collaboration 2003). In
this sense, this part of our criterion is rather similarly
defined. We require an
\begin{itemize}
\item {\it A priori protocol:} The populations that are to be
tested in the GLAST-LAT data must be defined before the initial
data release.
\end{itemize}

A {\it protocol} is technically the budget for testing
correlations. Every test will consume part of this budget up to a
point that, if we still proceed in testing, there can be no
statistical significant detection claim achieved anymore. A
protocol secures that a detection of a population can be made with
confidence in its statistical significance for a number of
interesting classes. As remarked by the Pierre Auger Collaboration
(2003), when confronted with claims made in the absence of an a
priori protocol, one may assume that a very large number of failed
trials were made in order to find the positive results being
reported, and thus disregard the claims altogether just by denying
statistical weight. Otherwise stated, we might be asked for proof
that the penalty for failed trials has been accounted for and is
indeed below a required statistical significance.\footnote{The
term {\it penalty} has an almost exact analog in humanities and
biology in the ``Bonferroni correction''.} This may turn out to
be, either very difficult to achieve or strictly impossible
because of the possible biases in penalties definitions.

Additional exploration of the same data set for expected or
unexpected populations can (and will) certainly be made, although
if the budget is spent, without the strength of immediate
discovery potential. A positive additional search must be thought
of as a way of pointing towards new populations of sources to be
tested with additional or independent sets of data
then.\footnote{A second year LAT catalog would not be independent:
it will combine already discovered sources of permanent character
with newly discovered ones that were below the sensitivity of 1 yr
integration. However, one possibility to have an indeed
independent sample with which produce future checks  would be
given by the blind keeping of a subsample of detections. (An
algorithm in the pipeline that would randomly choose say 1
$\gamma$-ray source in 10, and keep them out of all subsequent
population analysis and the reported catalog.) The unblinding of
this mini-catalog at a later stage would allow testing the most
promising of the populations that could not be confirmed in the
original sample because of the budget being completely spent. }

Summarizing, if using the same set of data, claiming the discovery
of one population affects the level of confidence by which one can
claim the discovery of a second. Then, suppose for definiteness
that the {\it total} budget is a chance probability equal to
${\cal B}$, e.g.,  10$^{-4}$. That is, that a claim for
population(s) discovery has to be better than one having a
probability of chance occurrence equal to ${\cal B}$, and that we
want to test $A$, $B$, $C$ \ldots classes of different sources
(say, radio galaxies, starburts galaxies, microquasars, pulsars,
AGN, etc.). The total budget can then be divided into individuals,
a priori, chance probabilities, $P_A$, $P_B$, etc., such that
$\sum_i P_i={\cal B}$.\footnote{The $\sum_i P_i$ also accounts for
any attempt to investigate population properties of subsamples
belonging to the same object class by invoking cuts. If too many
subsamples were investigated in order to discriminate further
among the emission characteristics in an already detected source
population, such selections are on the expense of the budget, too.
Statistically dependent test are to be avoided. A minimal set of
subsamples, imposing substantially different cuts in their
selections, is the most  adequate choice to maximize the chance
for statistically-significant classifications of subsamples.  }
This implies that population $i$ will be claimed as detected in
this framework if the a posteriori, factual, probability for its
random correlation, $P^{\rm LAT}(i)$, is less than the a priori
assigned $P_i$ (as opposed to be less only than the larger, total
budget).

We could go a step forward and manage the budget of probabilities.
For some populations, e.g., those which were not detected in EGRET
observations, we can less confidently assume that they will be
detected, or perhaps for some others, the number of their members
may be low enough such that a detection of only several of its
individuals would be needed to claim a large significance. In this
situation we would choose a relatively higher $P_i$, so that it
would be easier to find $P^{\rm LAT}(i) < P _i$. For others, say
AGN and pulsars, we know that they will be detected, and thus we
would be less willing to spent a large fraction of the discovery
budget in them. Within the protocol, we can statistically prove
that these population appear with very high confidence by
assigning a very low $P_i$ in such a way to make harder for the
test to pass. If one or more of the tests, i.e., if for several
$i$-classes, $P^{\rm LAT}(i) < P_i$, is fulfilled, the results are
individually significant. First, because we protected our search
by the a priori establishment of the protocol (a blind test) and
second, because the overall chance probability is still less than
the total budget ${\cal B}$.

We refrain ourselves here to explicitly propose which are the
populations to be tested and how large the a priori probability
assigned to each of them as well as the exact number for the total
budget ${\cal B}$ should be. This ultimately has to be carefully
studied by the LAT collaboration, although obvious choices can be
compiled and argued. We only emphasize that this should be done
before the data is taken. Now we proceed towards a most delicate
issue, that of the treatment of the statistical significance of
claimed detections of source populations.

\subsection{How to search and significance assessment}

The last constituent of a methodological approach to identify new
classes of $\gamma$-ray sources is the application of a

\begin{itemize}
\item {\it Common significance assessment:} we urge that a strict
statistical evaluation is mandatory before a claim of a discovery
of a new source population can be made. An objective method is
presented in the following.
\end{itemize}

We start by assessing the number of members of the relevant
candidate class being probed, for which predictions exist, that
coincide with LAT source detections of unidentified $\gamma$-ray
sources. Let ${\cal C}(A)$ represent this number for population
$A$.
In what follows, for the sake of simplicity, we will assume that
we deal with equally probable coincidences, when a projected
position is less distant than, say, the 95\% confidence contour.

Let ${\cal N}(A)$ be the number of known sources in the particular
candidate population $A$ under analysis and ${\cal U}$ the number
of LAT detections.
Let ${\cal P}$ be the probability that in a random direction of
the sky we find a LAT source.
The probability ${\cal P}$ should take into account instrumental
detectability issues (exposure gradients, imprecision of the
diffuse emission model, etc.) as well as, at low Galactic
latitudes, expected Galactic structures.

As an example which omits the latter complications, one may use
angular coverage (the ratio between the area covered by ${\cal U}$
sources and that of the sky region upon which these sources are
projected). In what follows, we will assume that such method is in
place for LAT and that ${\cal P}$ can be computed for a given
region of the sky. Note that to compute ${\cal P}$ we do not need
any information about the candidates, but just some sensible
extrapolation of the expected number of detections of sources that
have been already identified. The value of ${\cal P}$ is obtained
a priori of checking for any population.




Whatever the method, ${\cal P}$ is expected to be small for LAT.
To give an example, if we take just a coverage assessment at high
Galactic latitudes ($|b|>10$), and we assume that there will be a
thousand detections, and that the typical size of the error box of
LAT sources is a circle of radius 12 arcmin, then ${\cal P} \sim 3
\times 10^{-3} $. At lower latitudes, we expect ${\cal P}$ to be
between 1 to 2 orders or magnitude larger. We believe that a more
careful treatment of source number predictions and the range of
expected source location uncertainties will reduce the value of
${\cal P}$ from such simple estimations. Such low values for
${\cal P}$ make the product ${\cal P} \times {\cal N}(A)$ typical
less than 1-10, for all different candidate populations. We will
refer to this product as the {\it noise expectation}, i.e., this
is the number of coincidences which one would expect even when
there is no physical connection between the LAT detections and
population $A$.

The number of excess detections above noise will be, ${\cal
E}(A)={\cal C}(A)-{\cal P} \times {\cal N}(A)$.\footnote{
Trivially,  if the number of sources is so large that ${\cal P}
\rightarrow 1$, then ${\cal E}=0$. If instead, the number of
members in the potential counterpart class is so large that ${\cal
C}(A) \rightarrow {\cal P} {\cal N}(A)$, then ${\cal E}=0$ too. In
both cases, there is no way to distinguish whether the population
is physically associated. To simplify the treatment we consider
excesses with no overlapping, i.e., coincidences between members
of population A and LAT sources that are not co-spatial with
members of other populations. In reality, the available
$\gamma$-ray observables will allow further discrimination, either
directly by reducing overlap between members of different
populations at higher photon energies (better source localization
due to narrower instrumental psf), or when the populations under
consideration become distinguishable due to their source spectra,
and variability pattern.}
%
Two cases can be distinguished. The two largest populations of
plausible candidates (pulsars and blazars) will also present the
largest number of coincidences, since it is already proven that
they do emit high energy $\gamma$-rays above LAT sensitivity.
Let's assume that there are 2000 catalogued AGNs; with the quoted
value of ${\cal P}$, all coincidences in excess than 6 are beyond
the random expectation. The reality of the population in the EGRET
catalog make us expect that ${\cal C}({\rm AGN})\gg 6$, and thus
that the number of excesses would be equally large. In this case,
we are in the domain of large number statistics and a probability
for the number of excesses to occur by chance, $P^{\rm LAT}({\rm
AGN})$ can be readily computed.

A different case appears when the second term in the expression
for ${\cal E}(A)$ is a small quantity. Two scenarios may be found:
if the number of coincidences for that population is large
compared with the noise, we are again in the domain of large
number statistics, as in the case of AGN or pulsars. This will
--most likely-- not happen for many (or perhaps for any) of the
new populations we would like to test. Thus, in general we are in
the realm of small number statistics: we should test the null
hypothesis for a new source population against a reduced random
noise (see Feldman \& Cousins 1998, also Gehrels
1986).\footnote{If a precise number of detectable sources is
predicted, generally one could test the hypothesis of their
presence in the LAT catalog directly, using small number
statistics described in more detail below. However, this will not
constitute the standard scenario since we will not precisely from
theoretical arguments know how many, say, of the X-ray binaries,
should indeed be detectable. Modeling is in general not applied
with an equal level of detail to a sufficiently high number of
members in a candidate population.}

Let us analyze now an explicit example. We are testing a null
hypothesis (e.g., X-ray binaries are not LAT sources). That is
represented by 0 predicted signal events (coincidences), i.e.
total number of events equal to the background in Table 2-9 (see
leftmost columns) of Feldman \& Cousins (1998). Suppose for
definiteness that ${\cal P} \sim 3 \times 10^{-3}$ and ${\cal
N}(A)$ is equal to, say, 200, then the number of chance
coincidences (the noise or background) is 0.5. Thus, if we find
more than 5 individual members of this class (e.g. superseding the
confidence interval 0.00-4.64) correlated with LAT sources, we
have proven that the null hypothesis is ruled out at the the 95\%
CL.

Using the small number statistics formalism, we can convert the
level of confidence achieved for each population into the factual
probability, i.e., $P^{\rm LAT}({\rm X-ray\,bin})$. Subsequently,
by comparing with the a priori budgeted requirement (i.e., is
$P^{\rm LAT}({\rm X-ray\, bin.}) < P_{\rm X-ray\, bin.}$?, we will
be able to tell whether the population has been discovered.
Clearly, if instead we find no more than 5 individual sources in
the same example, then we have no evidence by which to claim the
existence of this population at that level of
confidence.\footnote{Notations: A confidence interval $[\mu_1,
\mu_2]$ is a member of a set, such that the set has the property
that $P(\mu \in [\mu_1, \mu_2]) = \alpha$, i.e., the probability
to find $\mu$ in the interval $[\mu_1, \mu_2]$ is $\alpha$. Here
$\mu_1$ and $\mu_2$ are functions of an observable $x$, and the
previous equation refers to the varying confidence intervals
$[\mu_1, \mu_2]$ from an ensemble of experiments with fixed $\mu$.
For a Poisson process, when the observable $x$ is the total number
of observed events $n$, consisting of signal events with expected
mean $\mu$ (in the case of a null hypothesis, $\mu=0$), and
background events with mean $b$, the probability of measuring $n$
given $\mu$ is $P(n|\mu) = (\mu + b)^n exp(-(\mu + b))/n!$ This
can be used to translate a given number of coincidences into a
probability for it to happen by chance.}

Managing $P_A$ is equivalent to requesting different populations
to appear with different, intelligently selected, levels of
confidence. By using this method, detecting just a few members of
each class may allow to achieve significant levels of confidence,
justified by the existence of the imposed theoretical censorship
and protected by an a priori protocol. Note that at this stage
there is no variability analysis involved. If we were to add the
search on compatible variability timescales, the confidence level
of the detections will even improve.

\section{Concluding remarks}

The proposed criterion for identification of $\gamma$-ray source
populations integrates three different parts:
1) A {\it theoretical censorship} that prohibits executing
repeated searches that would reduce the statistical significance
of any possible positive class correlation.
2) An {\it a priori protocol} that protects the significance by
which to claim the discovery of a number of important population
candidates and gives guidelines as to how to manage the
probability budget
3) A {\it significance assessment} that assigns probabilities both
in the large and in the small numbers statistical regime.


The potential of this methodological procedure is not limited to
the anticipated cases explicitly discussed here. By applying the
proposed scheme, one can also check spurious classifications in an
objective way, and test subsamples among the expected classes of
sources (e.g., FSRQs in correspondence of their peak radio flux,
or BL Lacs in correspondence of their peak synchrotron energy,
i.e. LBLs vs. HBLs, galaxy clusters in correspondence of their
X-ray brightness or ). Summarizing, the portrayed identification
scheme is not exclusively elaborated for source populations in
high-energy $\gamma$-rays. It's a methodological approach to be
generally applicable if the identification of source populations
among a complex astrophysical dataset can only be achieved by a
statistically sound discrimination between candidate classes.

\acknowledgments

The work of DFT was performed under the auspices of the U.S.
D.O.E. (NNSA), by the University of California Lawrence Livermore
National Laboratory under contract No. W-7405-Eng-48. OR
acknowledges support by DLR QV0002. We thank Luis Anchordoqui,
Seth Digel, and Dave Thompson for their valuable comments. The
remarks of an anonymous referee were used to improve this
manuscript.

\end{document}